\documentclass[twocolumn,superscriptaddress,prl]{revtex4}    
\usepackage{graphicx,amsmath,mathrsfs}
\usepackage{amssymb}
\usepackage{amsthm,multirow}
\usepackage{wasysym}  
\usepackage{subfigure,xcolor}   

\usepackage{color}

\def\bc{\begin{center}}
\def\ec{\end{center}}

\newcommand{\bs}[1]{\boldsymbol{#1}}

\def\ea{\emph{et al.}}

\begin{document}
\title{Three-band Hubbard model for Na$_2$IrO$_3$:\\
Topological insulator, zigzag antiferromagnet, and Kitaev-Heisenberg material}
\author{Manuel Laubach}
\affiliation{Institut f\"ur Theoretische Physik, Technische Universit\"at Dresden, 01062 Dresden, Germany}
\author{Johannes Reuther}
\affiliation{Dahlem Center for Complex Quantum Systems and Fachbereich Physik, FreieUniversit\"at Berlin, 14195 Berlin,\,Germany}
\affiliation{Helmholtz-Zentrum Berlin f\"ur Materialien und Energie, D-14109 Berlin, Germany}
\author{Ronny Thomale}
\affiliation{Institut f\"ur Theoretische Physik, Universit\"at W\"urzburg, 97074 W\"urzburg, Germany}
\author{Stephan Rachel}
\affiliation{Institut f\"ur Theoretische Physik, Technische Universit\"at Dresden, 01062 Dresden, Germany}
\affiliation{Department of Physics, Princeton University, Princeton, New Jersey 08544, USA}
\affiliation{School of Physics, University of Melbourne, Parkville, VIC 3010, Australia}

\begin{abstract}
Na$_2$IrO$_3$ was one of  the first materials proposed to feature the Kane-Mele type topological insulator phase. Contemporaneously it was claimed that the very same material is in a Mott insulating phase which is described by the Kitaev-Heisenberg (KH) model. First experiments indeed revealed Mott insulating behavior in conjunction with antiferromagnetic long-range order. Further refined experiments established antiferromagnetic order of zigzag type which is not captured by the KH model. Since then several extensions and modifications of the KH model were proposed in order to describe the experimental findings. Here we suggest that adding charge fluctuations to the KH model represents an alternative explanation of zigzag antiferromagnetism. Moreover, a phenomenological three-band Hubbard model unifies all the pieces of the puzzle: topological insulator physics for weak and KH model for strong electron--electron interactions as well as a zigzag antiferromagnet at 
intermediate interaction strength.
\end{abstract}


\maketitle


%

{\it Introduction.} The past decade of condensed matter physics has been strongly influenced by the rise of spin orbit coupling. Although known from the early days  of quantum mechanics, only recently it 
started to unfold its complexity and potential. Most notably, topological insulators\,\cite{hasan-10rmp3045,qi-11rmp1057} are spin orbit dominated band insulators with exotic surface properties. Amongst strongly correlated systems, spin-orbital liquids\,\cite{chen-09prl096406} and spin orbit--assisted Mott insulators\,\cite{kim-08prl076402} are fascinating states of matter which were proposed in the past years.

Soon after the theoretical prediction of the quantum spin Hall (QSH)\,\cite{kane-05prl226801} effect it was realized that graphene has negligibly small spin orbit coupling such that the topological insulator phase is not observable. Ever since, the field has been searching for other honeycomb lattice materials with possibly heavier elements and alternative mechanisms to accomplish large gap QSH states in honeycomb monolayers\,\cite{reis-16arXiv1608.00812}. 
One of the first proposals to feature Kane-Mele-type QSH physics focussed on Na$_2$IrO$_3$ where the Ir atoms form an effective 2D honeycomb net\,\cite{shitade-09prl256403}. Shitade \ea\ argued that if Coulomb interactions are not too strong Na$_2$IrO$_3$ might exhibit a correlated topological insulator (TI) ground state. Around the same time, Jackeli and Khaliullin proposed that Na$_2$IrO$_3$ might be a Mott insulator where the spin physics is dominated by both Heisenberg and exotic Kitaev spin exchange\,\cite{jackeli-09prl017205}, referred to as Kitaev-Heisenberg (KH) model. Since the pure Kitaev model\,\cite{kitaev06ap2} is a rare example of an exactly soluble quantum spin liquid system, the prospect of approximately realizing it in an actual material has attracted enormous attention. The phase diagram of the KH model contains a N\'eel ordered antiferromagnet, the Kitaev spin liquid, and a stripy antiferromagnet as an intermediate phase\,\cite{Chaloupka2010}.
The successful growth of single crystals, however, established a zigzag antiferromagnetic ground state\, \cite{choi-12prl127204,ye-12prb180403} -- as suggested earlier by {\it ab initio} calculations\,\cite{liu-11prb220403} -- which cannot be explained by the original KH model. 
Several works proposed extensions of this model\,\cite{kimchi-11prb180407,singh-12prl127203,mazin-12prl197201,chaloupka-13prl097204,foyevtsova-13prb035107,reuther-14prb100405,katukuri-14njp013056,Rau2014,yamaji-14prl107201,sizyuk-14prb155126,kimchi-15prb245134,rousochatzakis-15prx041035,winter-16prb214431} to account for the experimental findings, ranging from longer-ranged Heisenberg exchange\,\cite{kimchi-11prb180407,singh-12prl127203,winter-16prb214431} and
``negative'' KH interactions\,\cite{chaloupka-13prl097204}
 to off-diagonal $\Gamma$ exchange\,\cite{Rau2014,yamaji-14prl107201} and longer-ranged Kitaev interactions\,\cite{reuther-14prb100405,sizyuk-14prb155126,rousochatzakis-15prx041035}.
 Furthermore, the formation of molecular orbital crystals has been proposed as an explanation for the phenomenology of Na$_2$IrO$_3$\,\cite{mazin-12prl197201}. 
 Despite the variety of different models,
 it remains elusive until today which might be the spin Hamiltonian to most accurately describe the phenomenology of Na$_2$IrO$_3$.
Moreover, it has even been questioned whether Coulomb interactions are as strong as believed and whether the Mott limit is justified: {\it ab initio} results rather suggest a moderate interaction strength\,\cite{mazin-12prl197201,foyevtsova-13prb035107} or even a strong 3D TI phase\,\cite{kim-12prl106401}; likewise, recent ARPES experiments\,\cite{alidoust-16prb245132} report the observation of metallic surface states which are not compatible with the picture of an antiferromagnet in the {\it deep} Mott limit.

Given the enormous interest in the KH model and its extensions, one might wonder how a corresponding Hubbard model could be reconstructed of which the KH model is its strong coupling limit. A first guess might be that the non-interacting TI Hamiltonian considered by Shitade \ea\,\cite{shitade-09prl256403} leads in the strong coupling limit to the KH model. A simple analysis reveals, however, that in this case Kitaev spin exchange is generated on second neighbor links, while nearest neighbor exchange is of pure Heisenberg type\,\cite{reuther-12prb155127}. Likewise, one can construct a single-orbital band structure which indeed leads to the KH Hamiltonian for large $U$\,\cite{hassan-13prl037201}; it turns out, however, that such a band structure explicitly breaks time-reversal symmetry since real spin-dependent hoppings on nearest neighbor bonds are involved.

In the first part of this paper, we introduce a three-band Hubbard model with a time-reversal invariant band structure which is constructed such that its strong coupling limit is the pure KH spin model. The minimal model we identified has three orbital degrees of freedom which can be interpreted as the $t_{2g}$ manifold. Thus the phenomenological model introduced here has a strong similarity to the model studied earlier by Rau \ea\,\cite{Rau2014}. The analysis of the non-interacting band structure reveals an extended topological insulator phase. Since there are only two types of time-reversal invariant insulating bandstructures in 2D\,\cite{kane-05prl146802}, the weak-coupling limit of our three-band model is topologically equivalent to the one proposed by Shitade \ea\,\cite{shitade-09prl256403}. 
As we analyze the model in the presence of increasing interaction strength we find that it hosts a zigzag ordered phase at intermediate interaction strength $U$ in consistency with experiments for Na$_2$IrO$_3$.
In this sense, our phenomenological model
unifies the work by Shitade\,\ea\,\cite{shitade-09prl256403} at weak $U$, the work by Jackeli and Khaliullin\,\cite{jackeli-09prl017205} at strong $U$, and the experimental results\,\cite{choi-12prl127204,ye-12prb180403} at intermediate $U$.

In the second part, we choose the model's parameters to better describe Na$_2$IrO$_3$. We show that the zigzag phase at intermediate interaction strengths persists and we compute single particle spectral functions which can be compared to ARPES spectra. We demonstrate that topologically trivial edge states as remnants of the nearby topological insulator phase are present in the band gap of this three-band model.

{\it Phenomenological model.} The spinfull three-band model investigated in this work consists of a kinetic, spin-orbit and interaction term
\begin{equation}
H=H_\text{kin}+H_\text{SO}+H_\text{I}\label{ham}
\end{equation}
with 1/6 electron filling. The kinetic part takes the form
\begin{equation}
H_\text{kin}=\sum_{\langle i,j\rangle} \sum_{n,n',\alpha}d_{in\alpha}^\dagger T_{nn'}^{\gamma(i,j)}d_{in'\alpha}+\text{h.c.}\;,\label{hkin}
\end{equation}
where $d^{(\dagger)}_{in\alpha}$ denotes a fermionic annihilation (creation) operator on site $i$, with orbital $n\in\{yz,xz,xy\}$ and spin $\alpha\in\{\uparrow,\downarrow\}$. The orbital indices are reminiscent of the common labelling of a $t_{2g}$ manifold which our model is closely related to. The sum is over pairs of nearest neighbor sites $\langle i,j\rangle$ on the honeycomb lattice and $\gamma(i,j)\in\{x,y,z\}$ indicates the type of inequivalent Kitaev bond the pair $\langle i,j\rangle$ belongs to. For a $z$-bond the hopping matrix $T_{nn'}^z$ is given by
\begin{equation}
T^z=
\left(\begin{array}{ccc}
t_1 & t_2 & 0\\
t_2 & t_1 & 0\\
0 & 0 & t_1'
\end{array}\right)\;,\label{hoppingmatrix}
\end{equation} 
and the matrices $T_{nn'}^x$ and $T_{nn'}^y$ follow by cyclic permutations of the rows and columns. Unless explicitly stated otherwise we will assume $t_1=t_1'$ in the following. The spin-orbit coupling has the usual form 
\begin{equation}
H_\text{SO}=\frac{\lambda}{2}\sum_i \sum_{n,n',\alpha,\alpha'}d^\dagger_{in\alpha}\mathbf{L}_{nn'}\cdot\bs{\sigma}_{\alpha\alpha'}d_{in'\alpha'}\;,\label{hso}
\end{equation}
where $\mathbf{L}$ denotes an angular momentum operator in $3\times3$ matrix representation with $\mathbf{L}^2=l(l+1)=2$ and $\bs{\sigma}$ is the Pauli vector. When $\lambda>0$ the spin-orbit coupling generates a low energy $J=1/2$ doublet, representing the subspace in which the spin physics of the KH model takes place.

\begin{figure}[t!]
\centering
\includegraphics[width=0.45\textwidth]{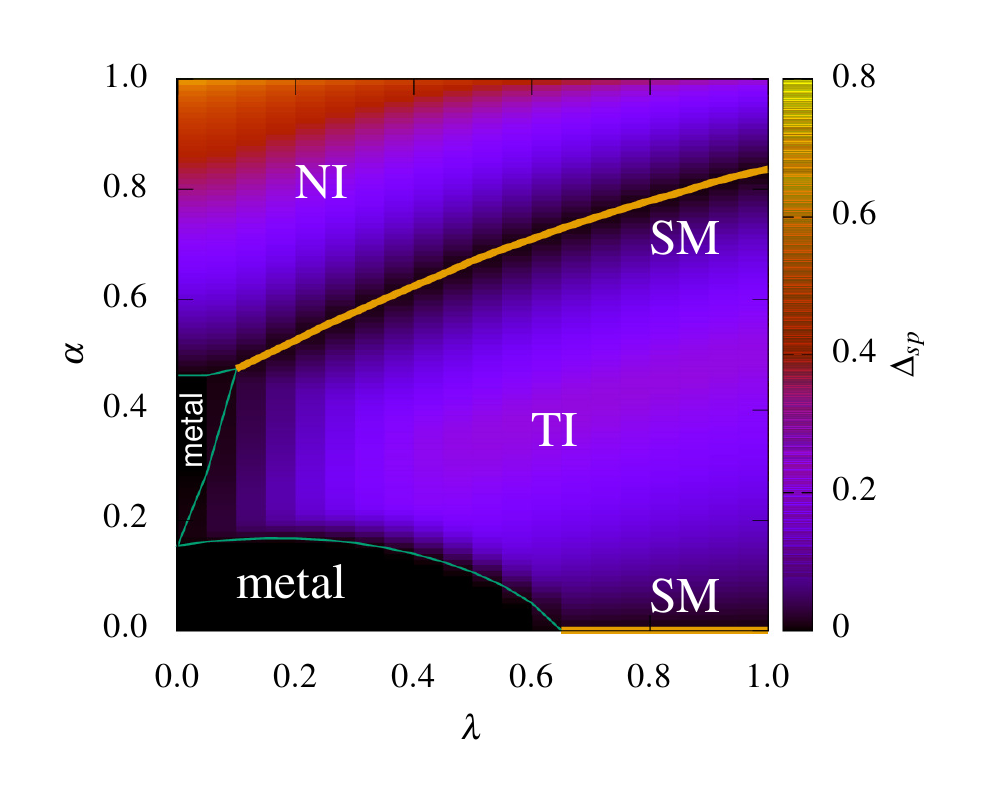}
\caption{Non-interacting $\lambda$-$\alpha$ phase diagram for $t_1'=t_1$ and $C=1$. The single particle gap $\Delta_{sp}$ is indicated via color code for the insulating phases. We find normal insulator (NI), topological insulator (TI), metallic phases, and semimetallic lines (SM).}
\label{fig:U=0}
\end{figure}

The simplest type of interaction term is an onsite Hubbard repulsion $H_\text{I}=\frac{U}{2}\sum_i (N_i-1)^2$ with $N_i=\sum_{n,\alpha} d^\dagger_{in\alpha}d_{in\alpha}$ which energetically favors single site occupancy. Performing a strong coupling expansion in second order in $t_{1/2}/U$ as described in Ref.~\cite{Rau2014} and projecting the result on the low energy $J=1/2$ subspace (which implies $U\gg t_{1/2}$ and $\lambda\gg t_{1/2}^2/U$), yield an usual spin-isotropic Heisenberg coupling $H=J\sum_{\langle i,j\rangle}\tilde{\mathbf{S}}_i\cdot\tilde{\mathbf{S}}_j$ with $J=\frac{4t_1^2}{U}$ where $\tilde{\mathbf{S}}_i$ are local spin-1/2 operators acting in the $J=1/2$ doublet. We find that the minimal SU(2) symmetric extension of the plain Hubbard repulsion which generates anisotropic spin interactions in this subspace has the form
\begin{equation}
H_\text{I}=\frac{U-3J_\text{H}}{2}\sum_{i}(N_i-1)^2-2J_\text{H}\sum_i \mathbf{S}_i^2
\end{equation}
with
\begin{equation}
\mathbf{S}_i=\frac{1}{2}\sum_{n,\alpha,\alpha'}d_{in\alpha}^\dagger\bs{\sigma}_{\alpha\alpha'}d_{in\alpha'}\;
\end{equation}
and the Hund's coupling $J_\text{H}$. The prefactors have been chosen to resemble the Kanamori Hamiltonian which has been used to describe multiplet interactions in $t_{2g}$ shells\, \cite{Rau2014,Georges2013,Medici2011}. (Note that the full Kanamori Hamiltonian also contains additional $\mathbf{L}_i^2$ terms.) Together with Eqs.~(\ref{hkin}) and (\ref{hso}) this type of minimal interaction will be investigated for different coupling strengths below.

At strong coupling (assuming $U\gg t_{1/2}$ and $\lambda\gg t_{1/2}^2/U$) the Hamiltonian (\ref{ham}) reduces to the KH model
\begin{equation}\label{hamHK}
H_\text{KH}= J\sum_{\langle i,j\rangle}\tilde{\mathbf{S}}_i\cdot\tilde{\mathbf{S}}_j-K \sum_{\langle i,j\rangle} \tilde{\mathbf{S}}^{\gamma(i,j)}_i\cdot\tilde{\mathbf{S}}^{\gamma(i,j)}_j
\end{equation}
with parameters 
\begin{equation}
J=\frac{4t_1^2(3U-4J_\text{H})}{3U(U-4J_\text{H})}\;,\quad K=\frac{16J_\text{H}t_2^2}{3U(U-4J_\text{H})}\;.\label{jk}
\end{equation}
When $U>4J_{\text{H}}$ the model features antiferromagnetic Heisenberg and ferromagnetic Kitaev interactions (i.e., $J,K>0$) as has been originally proposed by Chaloupka \ea\,\cite{Chaloupka2010}. Casting this Hamiltonian into the convenient form where the exchange interactions are written as $J=1-\alpha$, $K=2\alpha$, Eq.~(\ref{jk}) yields the hopping amplitudes
\begin{equation}
  t_1=C \sqrt{\frac{1-\alpha}{3 U/J_H -4}}\;,\quad t_2=C\sqrt{\frac{\alpha}{2}}
    \label{eq:t1t2ratio}
\end{equation}
where the parameter $C$ sets the energy scale of the kinetic Hamiltonian. In total, $\alpha$, $U$, $J_\text{H}$, $\lambda$, $C$ define a five parameter model, for which we will study the effects of charge fluctuations away from the strong coupling limit (note that one parameter can be absorbed into an overall energy scale). This three band system can be considered as a simplified (and particle-hole transformed) model for Na$_2$IrO$_3$ as it has been studied in Ref.~\cite{Rau2014}. A significant simplification comes from setting $t_1=t_1'$ in Eq.~(\ref{hoppingmatrix}). Given the orbital structure of the $t_{2g}$ states, $t_1<t_1'$ would certainly better account for the microscopic situation in Na$_2$IrO$_3$. However, this generates additional off-diagonal and symmetric $\Gamma$-exchange, which complicates the effective spin model and may drive spiral types of magnetic order. We, hence, mostly consider $t_1=t_1'$ but also discuss the qualitative changes of our results when $t_1<t_1'$.

We first focus on the properties of the non-interacting model, $U=J_H=0$. Instead of $t_{1/2}$ we use $\alpha$ and the spin orbit amplitude $\lambda$ as parameters (here we keep $U/J_H=5$ in order to be consistent with the discussion of the interacting case below). 
Calculating the band structure, we identify (semi-)metallic and insulating phases where the $\mathbb{Z}_2$ topological invariant 
reveals both topologically trivial and non-trivial insulating regimes. These findings are summarized in the non-interacting phase diagram in Fig.\,\ref{fig:U=0} where we also show the single-particle gap for the TI and normal insulator (NI) phases.
  Since there are only two types of time-reversal invariant insulating band structures in 2D (trivial and topological)\,\cite{kane-05prl146802} the TI phase is topologically equivalent to the one proposed by Shitade\,\ea\ Hence, together with the above results from a strong coupling expansion, our three-band model naturally unifies the proposals of Shitade\,\ea\ and Jackeli and Khaliullin.

{\it Zigzag antiferromagnetic phase.}~In general, Hubbard models in $d=2$ are not exactly soluble and one is restricted to approximations. Below, we will present numerical evidence that there is a phase transition from the topological insulator phase into a zigzag antiferromagnet (AFM) at intermediate interaction strength based on the Variational Cluster Approach (VCA). 
VCA is a quantum cluster approach very similar to Cluster Perturbation 
Theory\,\cite{gros-93prb418,senechal-02prb075129} 
where the interacting Green's function is computed on a small cluster which is then used to construct the full Green's function for the infinitely large system within a self-consistent, variational scheme. 
Here we use a $C$-shaped four-site cluster with three orbitals per site corresponding to an effective 12-site cluster. For details about the method we refer the interested reader to Refs.\,\cite{potthoff03epjb335,laubach-14prb165136,laubach-15prb041106,laubach-16prb241102}. 
%
The VCA method\,\cite{potthoff03epjb335,Potthoff2003b} has been successfully applied to study interacting topological phases\,\cite{yu-11prl010401,hassan-13prl037201,laubach-14prb165136,laubach-16prb241102} as well as magnetically ordered systems\,\cite{senechal-05prl156404,Yamada2013,laubach-14prb165136,Misumi2015,laubach-15prb041106}.


\begin{figure}[t!]
\centering
\includegraphics[width=0.46\textwidth]{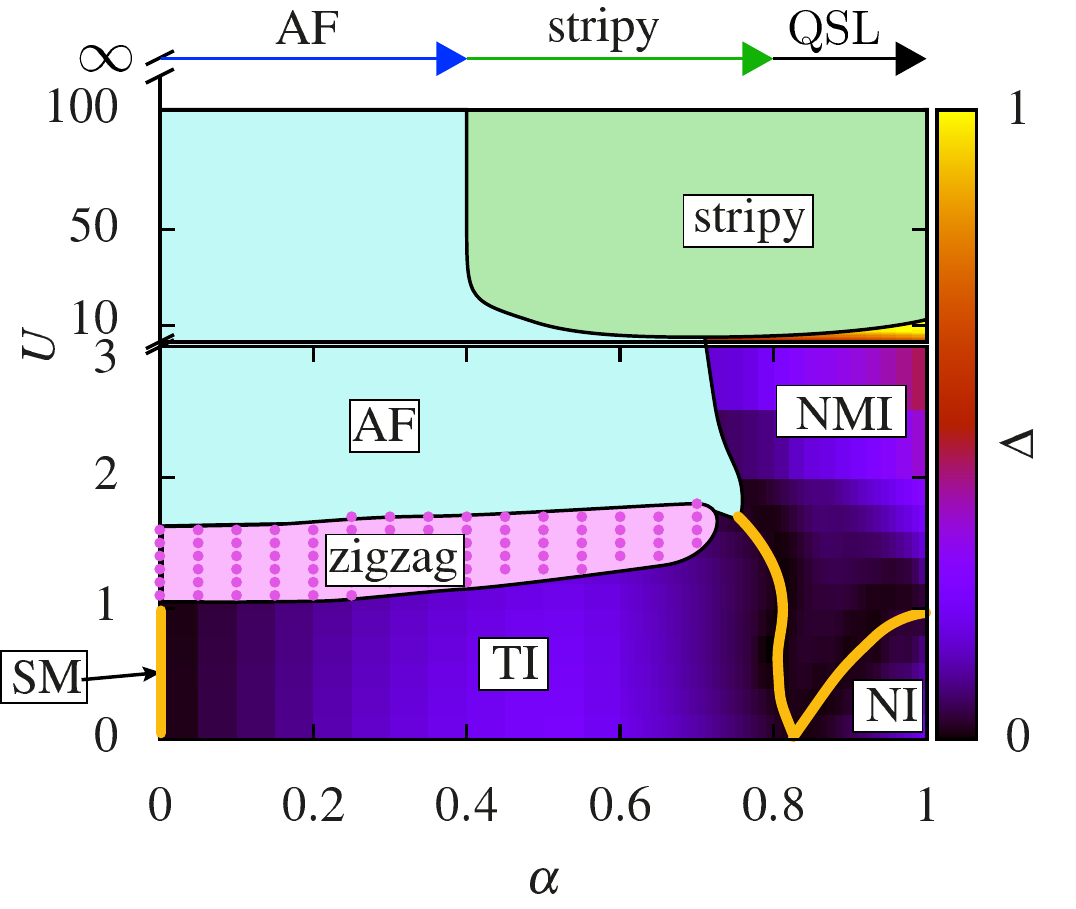} 
  \caption{
  Interacting phase diagram in the $U$-$\alpha$ plane for $\lambda=C=1$ and $t_1'=t_1$.
 Below the black horizontal line, the range from weak to intermediate interactions $0<U<3$ is shown, while above it, intermediate to strong interactions $3<U<50$. In addition, the spin limit $U\to\infty$ is shown. Finite $U$ results are obtained within VCA and the strong coupling spin model is solved using ED for 24 sites.
  We find topological (TI) and normal insulating (NI) phases, semi-metallic (SM) lines (brown), zigzag, N\'eel (AF), and stripy antiferromagnets, and a non-magnetic insulator (NMI) phase. For details see main text.
  %
}
\label{fig:phases}
\end{figure}

Using VCA, we obtain the $U$-$\alpha$ phase diagram for $0<U\leq 100$ and $\lambda=C=1$, see Fig.\,\ref{fig:phases}. To illustrate different coupling regimes, we divided the range of interactions into two parts: weak to intermediate interactions $0<U<3$ and intermediate to strong interactions $3<U<100$. In addition, we show the result in the strong coupling limit ($U\to\infty$) obtained from exact diagonalization (ED) for a finite system with 24 lattice sites which agrees with previous studies such as Ref.\,\cite{Rau2014}. As a central VCA result, we find that within wide ranges of $\alpha$, the topological insulator phase is stable up to $U\sim 1$ followed by a second order phase transition without gap closing into a zigzag AFM regime.

Mapping out the detailed interplay of different phases, we first observe a TI, NI and two semi-metallic points at $U=0$, see Fig.\,\ref{fig:phases}. Up to $U\sim 1$ the phase diagram remains largely unchanged except that the semi-metallic point between the TI and NI phases splits into two semi-metallic lines. For $U \apprge 1$ the TI phase, evolving over a wide range $0<\alpha<0.8$, undergoes a phase transition into the zigzag AFM phase. Note that the interaction driven phase transition from the TI phase into a collinear magnet is in general expected for non-frustrated lattices\,\cite{rachel-10prb075106,laubach-14prb165136}.
For $U\approx 1.6$ to $1.8$ a first-order phase transition from zigzag AFM to a N\'eel ordered AFM takes place. While the N\'eel phase extends up to $\alpha\approx0.8$ for small interactions, with increasing $U$ it slowly shrinks down to $\alpha\approx 0.4$ for $U=50$ which coincides with the phase transition in the ED phase diagram. For large $\alpha$ we detect a non-magnetic insulator (NMI) phase which persists up to $U\approx 10$ and down to very small $U$ at $\alpha\approx 0.85$. The origin of this phase is unclear: candidate states are quantum paramagnets including spin liquids and valence bond solids but also ordered states which are incommensurate with respect to the cluster size such as spiral magnets or large-unit cell magnets. At $U\approx 5$ a stripy AFM becomes stable for large $\alpha$ and starts to spread over an extended $\alpha$-range for increasing $U$. The reader should notice that VCA is unable to identify the Kitaev spin liquid phase, which is clearly resolved by ED for $\alpha>0.8$. Due to the small cluster size and the suppression of long-range entanglement (which is an inescapable consequence of quantum cluster approaches) this is by no means surprising. For large $U$ and $\alpha\approx1$, however, the VCA condensation energies of possible magnetic orders become almost indistinguishable. Such a nearly degenerate ground state scenario might indicate the proximity to a spin-liquid phase. Given the overall phase diagram in Fig.\,\ref{fig:phases}, a rough experimental estimate of $\frac{K}{J}\approx4$ (or even larger) as found in Ref.\,\cite{chun-15np462} would locate Na$_2$IrO$_3$ near the right boundary of the zigzag phase.

On a conceptually more simple level, the presence of a zigzag AFM at intermediate interaction strength can be further substantiated by a Hartree-Fock (HF) mean-field analysis. Recently, the HF approach has been applied to a three-band Hubbard model which is similar to Eq.\,\eqref{ham} but differs in details about the interaction terms and the signs of the parameters\,\cite{IgarashiNagao2015,IgarashiNagao2016}. In these works, a zigzag AFM phase was likewise identified at intermediate interaction strength. 


\begin{figure}[t!]
\centering

\hspace{-10pt}
\includegraphics[width=0.49\textwidth]{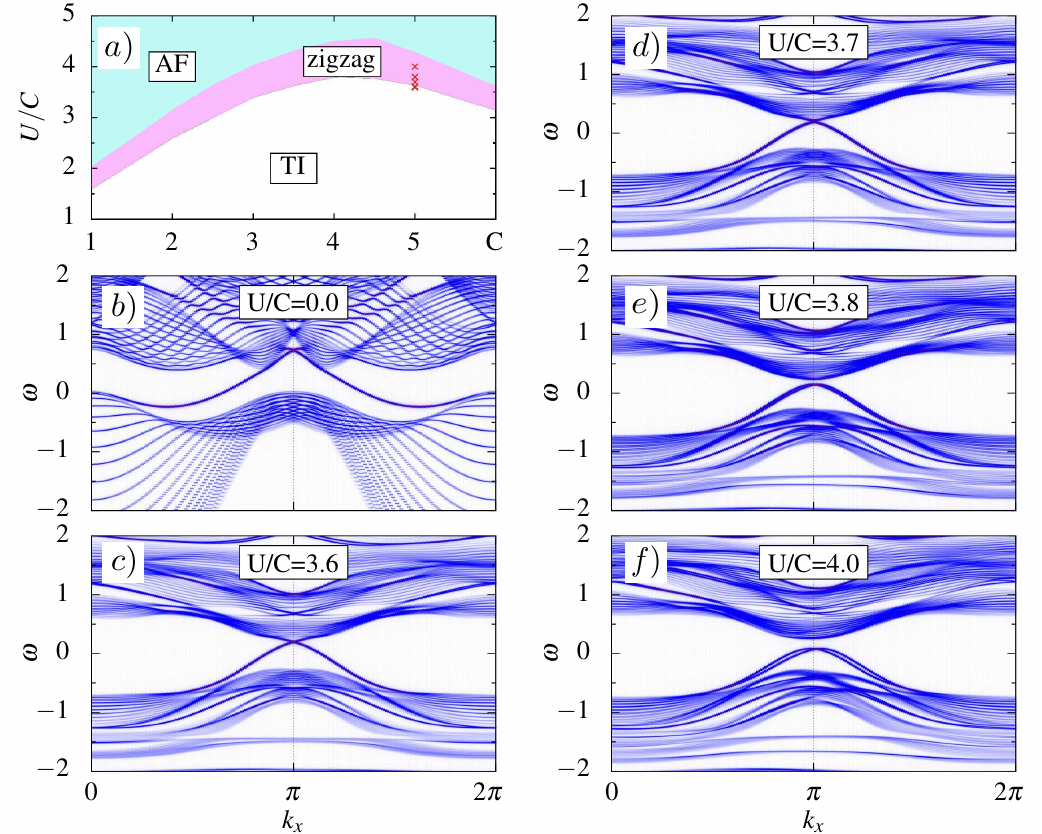} 
\caption{
(a) Phase diagram $U/C$ vs.\ $C$ for fixed $t_1' = 1.5 t_1$ and $\alpha=0.4$: the zigzag phase persists in the entire $C$ range but is shifted to larger values of $U/C$. (b-f) Spectral function $A(k_x,\omega)$ for the surface states at $t_1'=1.5t_1$, $\alpha=0.4$, $C=5$ and different interaction strength $U/C$. For $U/C\leq 3.6$ magnetic order is absent and helical edge states indicate the TI phase. 
For $U/C>3.6$ we find a finite zigzag order parameter and edge states acquire a small gap since the protecting time-reversal symmetry is spontaneously broken. 
The red crosses in panel (a) mark the position of the parameters used in (c-f).
}
\label{fig:spectral-zigzag}
\end{figure}

{\it Discussion.}~So far we considered $\lambda=C=1$ and $t_1 = t_1'$ where the latter condition led to the KH model in the strong coupling limit. 
The appropriate choice of effective parameters to account for the actual scenario found in experiment is a contentious issue and still under debate\,\cite{mazin-private}. From our findings, we can still make the general statement that zig zag magnetic order generically occurs in a large range of parameter space for intermediate Hubbard strength. 
Given the microscopic situation of Na$_2$IrO$_3$, the parameter choice $|\lambda|  \ll C$ and $t_1<t_1'$ certainly better accounts for the orbital structure and orientation of the $t_{2g}$ states as well as for {\it ab initio} results obtained earlier\,\cite{mazin-12prl197201,foyevtsova-13prb035107}. To make the model more realistic, we consider $t_1' = 1.5 t_1$, $\alpha=0.4$, and $0.8\leq C\leq6$ in the remainder of the paper [note that $C$ represents the overall kinetic energy scale, see Eq.~(\ref{eq:t1t2ratio})]. The consequences of this generalization are summarized as follows: (i) As mentioned before, due to $t_1' \neq t_1$ symmetric off-diagonal $\Gamma$ exchange is generated in the spin Hamiltonian \eqref{hamHK}. (ii) As a result of these couplings, the ED phase diagram at $U\rightarrow\infty$ shows an additional small 120$^\circ$ N\'eel regime between the stripy AFM phase and the Kitaev spin liquid\,\cite{Rau2014}. (iii) The non-interacting phase diagram only undergoes minor shifts of the phase boundaries. (iv) In the interacting VCA phase diagram, the zigzag phase persists in the whole range of investigated $C$ parameters but is shifted to larger values of $U/C$, see Fig.\,\ref{fig:spectral-zigzag}\,(a).

In the literature, there has been an intensive debate about what are the ``correct'' spin Hamiltonian and exchange couplings J, K, and $\Gamma$ for Na$_2$IrO$_3$\,\cite{jackeli-09prl017205,Chaloupka2010,choi-12prl127204,ye-12prb180403,liu-11prb220403,singh-12prl127203,mazin-12prl197201,chaloupka-13prl097204,foyevtsova-13prb035107,katukuri-14njp013056,Rau2014,yamaji-14prl107201,rousochatzakis-15prx041035,winter-16prb214431,reuther-14prb100405,sizyuk-14prb155126}. The couplings given in Eq.\,\eqref{jk}
will be modified for $t_1 \not= t_1'$ as follows,
\begin{eqnarray}
J&=& \frac{4}{9}\left( \frac{2 t_1^2 + {t_1'}^2}{U} + \frac{2 t_1(t_1 + 2 t_1')}{U - 4 J_H}\right) \\
K&=&-\frac{16 J}{9U} \frac{3 t_2^2 + ( t_1 - t_1')^2}{4J_H - U}\\[5pt]
\Gamma &=& \frac{32 J}{9U}\frac{t_2(t_1 - t_1')}{U - 4 J_H}
\end{eqnarray}
These expressions can be used to obtain sign and magnitude of the magnetic couplings depending on the input tight-binding parameters as well as $U$ and $J_H$. In principle, it can even be applied to the analogous discussions for other ``Kitaev-Heisenberg materials''\,\cite{trebst17arXiv1701.07056,hermanns-17arXiv:1705.01740,winter-17arXiv1706.06113}.

We finally focus on the regime near the transition between the TI and zigzag AFM phase. Assuming that the experimentally observed zigzag phase is indeed not deep in the Mott phase but much closer to the weak-coupling regime\,\cite{mazin-12prl197201,kim-12prl106401}
it is interesting to investigate whether remnants of the TI phase can still be detected in the zigzag phase. We compute the single-particle spectral function $A(k_x,\omega)$ on a ribbon geometry which allows to track the edge states even in the presence of interactions.
Using the above parameter setting $t_1' = 1.5 t_1$, $\alpha=0.4$ and fixing $C=5$, Fig.\,\ref{fig:spectral-zigzag} (b-f) shows the evolution of surface spectral functions for varying $U/C$. In the TI phase for $U/C\leq 3.6$ we clearly observe helical edge states traversing the bulk gap while the magnetization remains zero. At $U=3.7$ the second order phase transition occurs marking the smooth onset of magnetization. Once magnetic order sets in, we immediately observe a small gap in the edge states at $k_x=\pi$, which is a consequence of the lost topological protection. Even around $U/C=4$, we still observe the gapped edge states, see Fig.\,\ref{fig:spectral-zigzag}\,(f).
Interestingly, in a recent angle-resolved photo emission spectroscopy (ARPES) measurement on single crystals of Na$_2$IrO$_3$, metallic surface states with nearly linear dispersion have been revealed\,\cite{alidoust-16prb245132,catuneanu-16prb121118}. It remains as an exciting task to verify whether these metallic surface states are the topologically trivial remnants of the TI phase as suggested by our VCA analysis.


%


{\it Conclusion}.~We introduced a three-band Hubbard model which unifies three influential works in the context of Na$_2$IrO$_3$: the correlated TI phase proposed by Shitade \ea\,\cite{shitade-09prl256403}, the KH model advocated by Jackeli and Khaliullin\,\cite{jackeli-09prl017205}, and the inelastic neutron scattering results of Choi \ea\,\cite{choi-12prl127204} indicating a zigzag antiferromagnetic ground state. Our three-band Hubbard model hosts for weak $U$ an extended TI phase, at strong $U$ the KH model, and at intermediate $U$ a zigzag antiferromagnetic phase. Moreover, our analysis predicts topologically trivial edge states in the zigzag phase as a remnant of the TI phase which might have been observed in recent ARPES measurements\,\cite{alidoust-16prb245132}.


{\it Acknowledgements}.---
We acknowledge discussions with R.\ Valenti, N.\ Perkins, G.\ Jackeli, I.\ Mazin, and P.\ Gegenwart.
We acknowledge financial support
by the DFG through SFB 1143 (ML, SR) and SFB 1170 (RT),
by the ERC through the starting grant TOPOLECTRICS (ERC-StG-Thomale-336012, RT), and by the Freie Universit\"at Berlin within the Excellence Initiative of the
German Research Foundation (JR).
%
We thank the Center for Information Services and High Performance Computing (ZIH) at TU
Dresden for generous allocations of computer time.


\bibliographystyle{prsty}
\bibliography{iridate}

\end{document}